\documentclass[aps,showpacs]{revtex4}
\usepackage{graphicx}
\usepackage{dcolumn}
\usepackage{bm}
\newcommand{\car}{$^{12}$C }
\newcommand{\oxy}{$^{16}$O }
\newcommand{\caI}{$^{40}$Ca }

\newcommand{\bqu}{{\bf q}}

\newcommand{\cer}{{\cal R}}
\newcommand{\cde}{{\cal D}}
\begin{document}
\title {Superscaling in electroweak excitation of nuclei
       }
\author {M. Martini and G. Co'} 
\affiliation{Dipartimento di Fisica,  Universit\`a di Lecce,
and 
Istituto Nazionale di Fisica Nucleare  sez. di Lecce,
 I-73100 Lecce, Italy}
\author{M. Anguiano and A. M. Lallena}
\affiliation {Departamento de F\'{\i}sica
At\'omica, Molecular y Nuclear, Universidad de Granada, 
E-18071 Granada, Spain}

\date{\today}

\begin{abstract}
  Superscaling properties of \car, \oxy and \caI nuclear responses,
  induced by electron and neutrino scattering, are studied for
  momentum transfer values between 300 and 700 MeV/$c$.  We have
  defined two indexes to have quantitative estimates of the scaling
  quality.  We have analyzed experimental responses to get the
  empirical values of the two indexes.  We have then investigated the
  effects of finite dimensions, collective excitations, meson exchange
  currents, short-range correlations and final state interactions.
  These effects strongly modify the relativistic Fermi gas scaling
  functions, but they conserve the scaling properties.  We used the
  scaling functions to predict electron and neutrino cross sections and
  we tested their validity by comparing them with the cross sections
  obtained with a full calculation. For electron scattering we also
  made a comparison with data. We have calculated the total
  charge-exchange neutrino cross sections for neutrino energies up to
  300 MeV.
\end{abstract} 
\maketitle
%
\section{INTRODUCTION}
\label{sec:intro}

The properties of the Relativistic Fermi Gas (RFG) model of the
nucleus \cite{alb88,dep98,ama02} have inspired the idea of
superscaling.  In the RFG model, the responses of the system to an
external perturbation are related to a universal function of a
properly defined scaling variable which depends on the energy and
the momentum transferred to the system. With the adjective universal
we want to indicate that the scaling function is independent of the
momentum transfer, and also of the number of nucleons.  These
properties are respectively called scaling of first and second kind.
Furthermore, the scaling function can be defined in such a way to
result in independence also of the specific type of external one-body
operator.  This feature is usually called scaling of zeroth kind
\cite{mai02,ama05a,ama05b}.  One has superscaling when the three kinds
of scaling we have described are verified. This happens in the RFG
model.

The theoretical hypothesis of superscaling can be empirically tested
by extracting response functions from the experimental cross sections
and by studying their scaling behaviours. The responses of the nucleus
to electroweak probes can be extracted from the lepton-nucleus cross
sections by dividing them by the single-nucleon cross sections
properly weighted to account for the number of protons and neutrons.
In addition, one has to divide the obtained responses by the
appropriate electroweak form factors.

Inclusive electron scattering data in the quasi-elastic region have
been analyzed in this way \cite{don99a,don99b,mai02}. The main result
of these studies is that the longitudinal responses show superscaling
behaviour. To be more specific, scaling of second kind, independence
of the nucleus, is better fulfilled than the scaling of first kind,
independence of the momentum transfer. The situation for the
transverse responses is much more complicated.

The presence of superscaling features in the data is relevant not only
by itself, but mainly because this property can be used to make
predictions. In effect, from a specific set of longitudinal response
data \cite{jou96}, an empirical universal scaling function has been
extracted \cite{mai02} and has been used to obtain neutrino-nucleus
cross sections in the quasi-elastic region \cite{ama05a}.

We observed that this universal scaling function is quite different
from that predicted by the RFG model. This indicates the presence of
physical effects not included in the RFG model, but still conserving
the scaling properties. We have investigated the superscaling
behaviour of some of the effects not considered in the RFG model. They
are: the finite size of the system, its collective excitations, the
Short-Range Correlations (SRC), the Meson Exchange Currents (MEC) and
the Final State Interactions (FSI). The inclusion of these effects
produce scaling functions rather similar to the empirical ones.

Before presenting our results, we recall in Sec. \ref{sec:formalism}
the basic expressions of the superscaling formalism. We show how the
scaling functions are related to the electromagnetic and weak response
functions, and to the inclusive lepton scattering cross sections.

In Sec. \ref{sec:res1} we discuss the scaling properties of our
nuclear models.  To quantify the quality of the scaling between the
various functions obtained with different calculations, we define two
indexes, $\cer$ and $\cde$.  From the data of Ref. \cite{jou96} we
extract empirical reference values of these two indexes which indicate
if scaling has occurred.  From the same set of data we also extract an
empirical universal scaling function. We analyze the scaling
properties of all the effects beyond the RFG model, by comparing the
values of the two indexes $\cer$ and $\cde$ of the theoretical scaling
functions with the empirical ones. We choose a theoretical scaling
function obtained by including all the effects considered as a
theoretical universal scaling function.

In Sec. \ref{sec:res2} we study the prediction power of the
superscaling hypothesis.  Our universal empirical and theoretical
scaling functions are used to calculate electron and neutrino
inclusive cross sections. These results are compared with those
obtained by calculating the same cross sections with our nuclear
model. We discuss results for double differential electron scattering
processes, and compare our cross sections with experimental data. We
calculate also total neutrino cross sections for neutrino energies up
to 300 MeV.  In Sec. \ref{sec:conc} we summarize our results and
present our conclusions.

\section{BASIC SCALING FORMALISM}
\label{sec:formalism}
Scaling variables and functions have been presented in a number of
papers
\cite{alb88,bar98,don99a,don99b,mai02,bar04,ama05a,ama05b,ama06,ant06}.
The purpose of this section is to recall the expression of the scaling
variable used in our study and the relations between scaling
functions, responses and cross sections.
  
In this work we have considered only processes of inclusive lepton
scattering off nuclei. We have described these processes in Plane Wave
Born Approximation and we have neglected the terms related to the rest
masses of the leptons. In this presentation we indicate respectively
with $\omega$ and $\bqu$ the energy and the momentum transferred to
the nucleus.

In the RFG model the scaling variables and functions are related to
the two free parameters of the model: the Fermi momentum $k_{\rm F}$ and the
energy shift $E_{\rm shift}$. We define the quantity:
\begin{equation} 
\Psi_0 = \frac {2 m} {k_{\rm F}} 
\left[
\sqrt{\left( \frac{\omega -E_{\rm shift}}{2m}  \right) 
      \left(1 + \frac{\omega -E_{\rm shift}}{2m} \right) }
    - \frac{q}{2m}
\right] \, ,
\label{eq:psiz}
\end{equation}
where $q \equiv |\bqu|$, and $m=(m_p+m_n)/2$ indicates the average
nucleon mass. The scaling variable is then defined as:
\begin{equation} 
\Psi = \Psi_0 \left(
1+ \Psi_0 \frac{k_{\rm F}}{2q}\sqrt{\frac{q^2}{m^2}+1}
\,\right) \, .
\label{eq:psi}
\end{equation}
The RFG model provides a universal scaling function
which 
can be
expressed in terms of the scaling variable $\Psi$ as
\cite{ama05a,ama05b}: 
\begin{equation}
f^{\rm RFG}(\Psi) =
 \frac {3}{4} \left( 1-\Psi^2 \right)
          \theta\left(1-\Psi^2 \right)  \, ,
\label{eq:scalFG}
\end{equation}
where $\theta(x)$ indicates the step function. The RFG scaling
function (\ref{eq:scalFG}) is normalized to unity.

In the electron scattering case, the inclusive double differential
cross section can be written as \cite{bof96}:
\begin{equation} 
\frac{{\rm d}^2 \sigma}{{\rm d}\theta\,{\rm d}\omega} =
\sigma_{\rm M}
\left\{\frac{(\omega^2 - q^2)^2}{q^4} R_{\rm L}(\omega,q) + 
     \left[\tan^2 \left(\frac{\theta}{2}\right) 
      - \frac{\omega^2 - q^2}{2q^2}\right] R_{\rm T}(\omega,q)
\right\} \, ,
\end{equation}
where $\theta$ is the scattering angle, $\sigma_{\rm M}$ is the Mott
cross section, and we have indicated with $R_{\rm L}$ and $R_{\rm T}$
the longitudinal and transverse responses, respectively defined as:
\begin{equation}
R_{\rm L}(\omega,q)
=4 \pi \sum_{J=0}~|\langle J_{\rm f}||\mathcal{C}_J||J_{\rm i}\rangle |^2
\, ,
\label{eq:rl}
\end{equation}
and
\begin{equation}
R_{\rm T}(\omega,q)=
4 \pi \sum_{J=1}~(|\langle J_{\rm f}||\mathcal{E}_J||J_{\rm i}\rangle |^2+
|\langle J_{\rm f}||\mathcal{M}_J||J_{\rm i}\rangle |^2) \, .
\label{eq:rt}
\end{equation}
In the above equations we have indicated with $|J_{\rm i}\rangle$ and 
$|J_{\rm f}\rangle$ the initial and final states of the nucleus
characterized by their total angular momenta $J_{\rm i}$ and $J_{\rm f}$. 
The double bars indicate that the angular momentum matrix elements are
evaluated in their reduced expressions, as given by the Wigner-Eckart
theorem \cite{edm57}. We have indicated with $\mathcal{C}_J$,
$\mathcal{E}_J$ and $\mathcal{M}_J$ respectively the Coulomb, electric
and magnetic  multipole operators \cite{bof96,bla52}.

The scaling functions are obtained from the electromagnetic responses
as:

\begin{eqnarray}
f_{\rm L}(\Psi) &=& k_{\rm F} \, \frac{q^2 - \omega^2}{q\,m} \,
\frac {R_{\rm L}(\omega,q)}{Z (G^p_{\rm E})^2 + N (G^n_{\rm E})^2} \, ,
\label{eq:lscal}
\\
f_{\rm T}(\Psi) &=& 2\,k_{\rm F}\,  \frac{q\,m}{q^2 - \omega^2} \,
\frac {R_{\rm T}(\omega,q)}{Z (G^p_{\rm M})^2 + N (G^n_{\rm M})^2} \, ,
\label{eq:tscal}
\end{eqnarray}
where $Z$ and $N$ indicate, respectively, the number of protons and
neutrons of the target nucleus, and we have indicated with $G_{\rm
E,M}^{p,n}$ the electric (E) and magnetic (M) form factors of the
proton ($p$) and the neutron ($n$). In our calculations we used the
electromagnetic nucleon form factors of Ref. \cite{hoe76}. In
Eq. (\ref{eq:tscal}) only the magnetic nucleon form factors are
present. This implies the hypothesis that in $R_{\rm T}$ only the
one-body magnetization current is active. In the range of momentum
transfer values investigated, from 300 to 700 MeV/$c$, we found that the
contribution of the convection current is of few percents that of the
magnetization current.

Since our calculations are done in a non relativistic framework, we
have modified our responses by using the semi-relativistic corrections
proposed in \cite{ama05b,ama96}:
\begin{eqnarray}
\epsilon &\rightarrow& \epsilon 
\left(1 + \frac{\epsilon}{2 m} \right) \, ,
\label{eq:rele} \\
R_{\rm L}(q,\omega)  &\rightarrow&
\frac {q^2}{q^2-\omega^2} \, R_{\rm L}(q,\omega) \, , 
\label{eq:relrl} \\
R_{\rm T}(q,\omega)  &\rightarrow&
\frac{q^2-\omega^2}{q^2} \, R_{\rm T}(q,\omega) \, ,
\label{eq:relrt}
\end{eqnarray}
where $\epsilon$ indicates the energy of the emitted nucleon.

The above discussion can be extended to the case of the inclusive 
neutrino scattering processes. For example, for the $(\nu_e,e^-)$
reaction we express the differential cross section 
as \cite{oco72,wal75,wal95}:
\begin{eqnarray}
\nonumber
\frac{{\rm d}^2 \sigma}{{\rm d}\Omega {\rm d}\omega}&=&
\frac{G^2 \cos^2 \theta_{\rm C}}{(2 \pi)^2}| \,
{\bf k}_{\rm f}| \, \epsilon_{\rm f} \, F(Z',\epsilon_{\rm f})
\left\{\left(l_0 l_0^{\star}+\frac{\omega^2}{q^2}l_3 l_3^{\star}
-\frac{\omega}{q}l_3 l_0^{\star}\right) \, R_{\rm CC}^{\rm V}(\omega,q)
\right.
\\
\nonumber
&& + \, l_0 l_0^{\star} \, R_{\rm CC}^{\rm A}(\omega,q) \, + \,
l_3 l_3^{\star}  \,  R_{\rm LL}^{\rm A}(\omega,q) \,
+ \, 2 \, l_3 l_0^{\star} \, R_{\rm CL}^{\rm A}(\omega,q) \\
&& + \, \left.
\frac{1}{2} \, 
\left( \vec{l}\cdot\vec{l}^{\star}-l_3 l_3^{\star} \right) \,
\left[R_{\rm T}^{\rm V}(\omega,q)+R_{\rm T}^{\rm A}(\omega,q)\right] 
\, - \, 
\frac{i}{2} \, \left( \vec{l}\times\vec{l}^{\star} \right)_3 \,
R^{VA}_{\rm T'}(\omega,q)\right\} \, .
\label{eq:nuxsect}
\end{eqnarray}
In the above equation we have indicated with $G$ the Fermi constant,
with $\theta_{\rm C}$ the Cabibbo angle, with $\epsilon_{\rm f}$ and
${\bf k}_{\rm f}$ the energy and the momentum of the scattered lepton
and with $F(Z',\epsilon_{\rm f})$ the Fermi function taking into
account the distortion of the electron wave function due to the
Coulomb field of the daughter nucleus of charge $Z'$.  The expressions
of the factors $l_i l_i^{\star}$, related only to the leptons  
variables, are given in Refs. \cite{oco72,wal75,wal95}.

The nuclear response functions are expressed in terms of multipole
expansion of the operators describing the various terms of the weak
interaction. They are the
Coulomb $\mathcal{C}_J$, longitudinal
$\mathcal{L}_J$, transverse electric $\mathcal{E}_J$ and transverse
magnetic $\mathcal{M}_J$ operators. The responses are defined as: 
\begin{equation}
R_{\rm CC}^{\rm V}(\omega,q)=
4 \pi \sum_{J=0}~|\langle J_{\rm f}||
\mathcal{C}_J^{\rm V}||J_{\rm i}\rangle |^2 \, ,
\label{eq:rccv}
\end{equation}
\begin{equation}
R_{\rm CC}^{\rm A}(\omega,q)=
4 \pi \sum_{J=0}~|\langle J_{\rm f}||
\mathcal{C}_J^{\rm A}||J_{\rm i}\rangle |^2 \, ,
\label{eq:rcca}
\end{equation}
\begin{equation}
R_{\rm CL}^{\rm A}(\omega,q)=
2 \pi ~\sum_{J=0} \left(
\langle J_{\rm f}||\mathcal{C}_J^{\rm A}||J_{\rm i}\rangle^{\star}
\langle J_{\rm f}||\mathcal{L}_J^{\rm A}||J_{\rm i}\rangle+
\langle J_{\rm f}||\mathcal{C}_J^{\rm A}||J_{\rm i}\rangle
\langle J_{\rm f}||\mathcal{L}_J^{\rm A}||J_{\rm i}\rangle^{\star}
\right) \, ,
\label{eq:rcla}
\end{equation}
\begin{equation}
R_{\rm LL}^{\rm A}(\omega,q)=
4 \pi \sum_{J=0}~
|\langle J_{\rm f}||\mathcal{L}_J^{\rm A}||J_{\rm i}\rangle |^2 \, ,
\label{eq:rlla}
\end{equation}
\begin{equation}
R_{\rm T}^{\rm V}(\omega,q)
=4 \pi \sum_{J=1}\, \left(
|\langle J_{\rm f}||\mathcal{E}_J^{\rm V}||J_{\rm i}\rangle |^2+
|\langle J_{\rm f}||\mathcal{M}_J^{\rm V}||J_{\rm i}\rangle |^2
\right) \, ,
\label{eq:rtv}
\end{equation}
\begin{equation}
R_{\rm T}^{\rm A}(\omega,q)
=4 \pi \sum_{J=1} \, \left(
|\langle J_{\rm f}||\mathcal{E}_J^{\rm A}||J_{\rm i}\rangle |^2+
|\langle J_{\rm f}||\mathcal{M}_J^{\rm A}||J_{\rm i}\rangle |^2
\right) \, ,
\label{eq:rta}
\end{equation}
and
\begin{eqnarray}
R_{\rm T'}^{\rm VA}(\omega,q)
= 2 \pi~ \sum_{J=1}&&  \left(
\langle J_{\rm f}||\mathcal{E}_J^{\rm V}||J_{\rm i}\rangle^{\star}
\langle J_{\rm f}||\mathcal{M}_J^{\rm A}||J_{\rm i}\rangle +
\langle J_{\rm f}||\mathcal{E}_J^{\rm V}||J_{\rm i}\rangle
\langle J_{\rm f}||\mathcal{M}_J^{\rm A}||J_{\rm i}\rangle^{\star}+
\nonumber \right. \\ 
&&
\left. 
\langle J_{\rm f}||\mathcal{E}_J^{\rm A}||J_{\rm i}\rangle^{\star}
\langle J_{\rm f}||\mathcal{M}_J^{\rm V}||J_{\rm i}\rangle+
\langle J_{\rm f}||\mathcal{E}_J^{\rm A}||J_{\rm i}\rangle
\langle J_{\rm f}||\mathcal{M}_J^{\rm V}||J_{\rm i}\rangle^{\star}
\right) \, ,
\label{eq:rtpva}
\end{eqnarray}
where we have separated the vector (V) and the axial-vector (A)
contributions. 

We found that the terms related to the axial-Coulomb operator
$\mathcal{C}^{\rm A}_J$ give a very small contribution to the cross
section, and, in our study, we neglected them.  This means that our
scaling analysis has been done for the $R_{\rm CC}^{\rm V}$, $R_{\rm
  LL}^{\rm A}$, $R_{\rm T}^{\rm V}$, $R_{\rm T}^{\rm A}$ and $R_{\rm
  T'}^{\rm VA}$ responses only.  The corresponding scaling functions
have been defined as:
\begin{eqnarray}
f_{\rm CC}^{\rm V}(\Psi) &=& k_{\rm F}\, \frac{q^2 - \omega^2}{q\,m}
\, \frac {R_{\rm CC}^{\rm V}(\omega,q)}{N (G^{(1)}_{\rm E})^2} \, ,
\label{eq:fccv}
\\
f_{\rm LL}^{\rm A}(\Psi) &=& 4\,k_{\rm F} \,
\frac{q\,m}{4m^2+q^2 - \omega^2} \,
\frac {R_{\rm LL}^{\rm A}(\omega,q)}{N (G_{\rm A})^2} \, ,
\label{eq:flla}
\\
f_{\rm T}^{\rm V}(\Psi) &=& 2\,k_{\rm F} \, 
\frac{q\,m}{q^2 - \omega^2} \,
\frac {R_{\rm T}^{\rm V}(\omega,q)}{N (G^{(1)}_{\rm M})^2} \, ,
\label{eq:ftv}
\\
f_{\rm T}^{\rm A}(\Psi) &=& 2\,k_{\rm F} \, 
\frac{q\,m}{4m^2+q^2 - \omega^2} \,
\frac {R_{\rm T}^{\rm A}(\omega,q)}{N (G_{\rm A})^2} \, ,
\label{eq:fta}
\\
f_{\rm T'}^{\rm VA}(\Psi) &=& 2\,k_{\rm F} \, 
\frac{q\,m}{\sqrt{q^2 - \omega^2}\sqrt{4m^2+q^2 - \omega^2}} \,
\frac {R_{\rm T'}^{\rm VA}(\omega,q)}{N G^{(1)}_{\rm M} G_{\rm A}}  
\, ,
\label{eq:ftpva}
\end{eqnarray}
where we have indicated with $G^{(1)}_{\rm E,M}$ the isovector
electric (E) and magnetic (M) nucleon form factors, and with $G_{\rm
  A}$ the axial-vector one. We have used the electromagnetic form
factors of Ref. \cite{hoe76}, and the dipole form of the axial vector
form factor with an axial mass value of 1030 MeV.

The relativistic effects are taken into account by using
semi-relativistic corrections similar to those of Eqs.
(\ref{eq:rele})-(\ref {eq:relrt}). In this case
the responses are obtained by doing the following changes with respect
to the pure non relativistic case:
\begin{eqnarray}
R_{\rm CC}^{\rm V}(q,\omega)  &\rightarrow& 
\frac {q^2}{q^2-\omega^2}\,  R_{\rm CC}^{\rm V}(q,\omega)  \, ,
\label{eq:relrccv} \\
R_{\rm LL}^{\rm A}(q,\omega)  &\rightarrow&
\left(1+\frac{q^2-\omega^2}{4m^2}\right)\,  
R_{\rm LL}^{\rm A}(q,\omega) \, ,
\label{eq:relrlla} \\
R_{\rm T}^{\rm V}(q,\omega)  &\rightarrow& 
\frac{q^2-\omega^2} {q^2}\, R_{\rm T}^{\rm V}(q,\omega)  \, ,
\label{eq:relrtv}\\
R_{\rm T}^{\rm A}(q,\omega)  &\rightarrow&
\left(1+\frac{q^2-\omega^2}{4m^2}\right)\,  
R_{\rm T}^{\rm A}(q,\omega) \, ,
\label{eq:relrta} \\
R_{\rm T'}^{\rm VA}(q,\omega)  &\rightarrow& 
\sqrt{\frac {q^2-\omega^2}{q^2}}\,  
\sqrt{1+\frac{q^2-\omega^2}{4m^2}}\,  
R_{\rm T'}^{\rm VA}(q,\omega) \, .
\label{eq:relrtpva}
\end{eqnarray}

The extension of these expressions to antineutrino charge exchange
scattering processes is straightforward. The treatment of charge
conserving processes is slightly different.

\section{SUPERSCALING BEYOND RFG MODEL}
\label{sec:res1}
The basic quantities calculated in our work are the electromagnetic,
and the weak, nuclear response functions. We obtain the scaling
functions by using Eqs. (\ref{eq:lscal}) and (\ref{eq:tscal}) for the
electron scattering case, and Eqs. (\ref{eq:fccv})-(\ref{eq:ftpva})
for the neutrino scattering. The scaling properties of the scaling
functions have been studied by a direct numerical comparison. We
thought it necessary to define some numerical index able to quantify the
quality of the scaling.

Let us consider the problem of comparing a number $M$ of scaling
functions $\left\{{f_\alpha, \alpha=1,\ldots,M} \right\}$, each of
them known for $K$ values of the scaling variable $\left\{ \Psi_i,
i=1,\ldots,K \right\}$. For each value of $\Psi_i$ we found the
maximum and minimum of the various $f_\alpha$ scaling functions:
\begin{equation}
f^{\rm max}_i \, = \, 
\max_{\alpha=1,\ldots,M} \left[ f_\alpha(\Psi_i) \right] \, ,
\end{equation}
\begin{equation}
f^{\rm min}_i \, = \, 
\min_{\alpha=1,\ldots,M} \left[ f_\alpha(\Psi_i) \right] \, .
\end{equation}
We define the two indexes
\begin{equation}
\cde \, = \, \max_{i=1,\ldots,K} 
\left[ f^{\rm max}_i \, - \, f^{\rm min}_i \right] \, ,
\label{eq:delta} 
\end{equation}
and
\begin{equation}
{\cal R}=\frac {1}{K f^{\max}} \sum_{i=1,\ldots,K} 
\left[ f^{\rm max}_i - f^{\rm min}_i \right] \, ,
\label{eq:erre}
\end{equation}
where $f^{\rm max}$ is:
\begin{equation}
f^{\rm max} \, = \, 
\max_{i=1,\ldots,K} \left[ f^{\rm max}_i \right] \, .
\end{equation}
The two indexes give complementary information.  The $\cde$ index is
related to a local property of the functions: the maximum distance
between the various curves. The value of this index could be
misleading if the responses have sharp resonances. For this reason we
have also used the $\cer$ index which is instead sensitive to global
properties of the differences between the functions. Since we know
that the functions we want to compare are roughly bell shaped, we have
inserted the factor $1/f^{\rm max}$ to weight more the region of the
maxima of the functions than that of the tails.

The perfect scaling is obtained when both $\cde$ and $\cer$ are
zero. This is achieved only in the RFG model. In our calculations the
perfect scaling is obviously violated, as it is violated by the
empirical scaling functions.  In order to have reference values of the
two indexes we have determined the values of $\cer$ and $\cde$ for
experimental scaling functions extracted from the longitudinal and
transverse electromagnetic response data of $^{12}$C, \caI and $^{56}$Fe
given in Ref. \cite{jou96}. This is the same set of data used in
Ref. \cite{mai02} to extract a universal scaling function.

The definition of the scaling variable $\Psi$, Eqs. (\ref{eq:psiz})
and (\ref{eq:psi}), requires on to fix the values of $k_{\rm F}$ and
$E_{\rm shift}$ for each nucleus. We used values of $k_{\rm F}$
obtained by doing an average over the nuclear density, and values of
$E_{\rm shift}$ that, in a Fermi gas calculation, reproduce the peak
position of the experimental response functions \cite{ama94b}. We used
$E_{\rm shift}$=15 MeV for all the nuclei and $k_{\rm F}$=215 MeV/$c$
for \car and \oxy and $k_{\rm F}$=245 MeV/$c$ for \caI and $^{56}$Fe.

\begin{table}[b]
\begin{center}
\begin{tabular}{|c|c|c|c|c|}
\hline
          & \multicolumn{2}{c|}{$f_{\rm L}$} &  
            \multicolumn{2}{c|} {$f_{\rm T}$} \\
\hline 
$q$ [MeV/$c$] & $\cde$ & $\cer$ & $\cde$  &  $\cer$ \\
\hline
   300    & 0.107 $\pm$ 0.002 & 0.152 $ \pm$ 0.013 & 
            0.223 $\pm$ 0.004 &  0.165 $\pm$ 0.017  \\
   380    & 0.079 $\pm$ 0.003 & 0.075 $ \pm$ 0.009 & 
            0.235 $\pm$ 0.005 &  0.155 $\pm$ 0.014    \\
   570    & {\bf 0.101 $\pm$ 0.009} & {\bf 0.079 $ \pm$ 0.017} & 
            0.169 $\pm$ 0.003 &  0.082 $\pm$ 0.007  \\
\hline
\end{tabular}
\end{center}
\vspace*{-0.6cm}
\caption{\small Values of the $\cde$ and $\cer$ indexes,
  Eqs. (\ref{eq:delta}) and (\ref{eq:erre}), calculated by comparing
  the empirical $f_{\rm L}$ and $f_{\rm T}$ scaling functions shown in
  Fig. \protect\ref{fig:fexp} for each value of the momentum transfer
  $q$. The values provide information about the scaling of second
  kind. The values of the $\cde$ and $\cer$ for $f_{\rm L}$ at $q$=570
  MeV/$c$, in boldface, are taken as our reference values.  }
\label{tab:rdelta}
\end{table}

The details of the procedures we have used to extract the experimental
scaling functions and to calculate the empirical values of $\cer$ and
$\cde$ are given in Appendix \ref{sec:expdata}. We present in Fig.
\ref{fig:fexp} the experimental longitudinal and transverse scaling
function data for $^{12}$C, \caI and $^{56}$Fe for each value of the
momentum transfer given in Ref.  \cite{jou96}.  In Table
\ref{tab:rdelta} we give the values of $\cde$ and $\cer$ obtained by
comparing the experimental scaling functions shown in each panel.

We analyze the empirical scaling functions by studying the three
different kinds of scaling defined in the Introduction. The
presentation of the data of Fig. \ref{fig:fexp} and Table
\ref{tab:rdelta} gives direct information on the scaling of second
kind.  It is immediate to observe that, in this case, the $f_{\rm L}$
functions scale better than the $f_{\rm T}$ ones.  The $f_{\rm T}$
scaling functions of $^{12}$C, especially for the lower $q$ values, are
remarkably different from those of \caI and $^{56}$Fe. This is
confirmed by the values of $\cer$ and $\cde$ given in
Table \ref{tab:rdelta}.

\begin{figure}
\begin{center}
\parbox[c]{16cm}
{\includegraphics[scale=0.5]{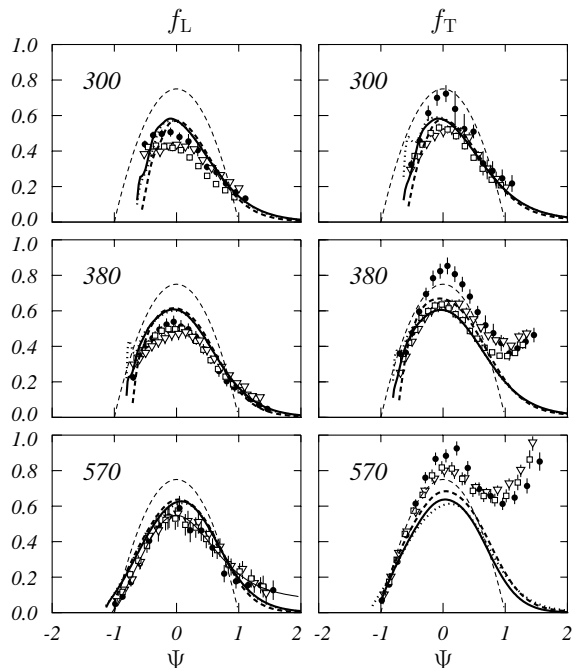}}
\end{center}
\vspace*{-0.8cm}
\caption{\small Empirical longitudinal, $f_{\rm L}$, and transverse,
  $f_{\rm T}$, scaling functions obtained from the experimental
  electromagnetic responses of Ref.  \protect\cite{jou96} as explained
  in Appendix \protect\ref{sec:expdata}.  The numbers in the panels
  indicate the values of the momentum transfer in MeV/$c$. The full
  circles refer to $^{12}$C, the white squares to $^{40}$Ca, and the
  white triangles to $^{56}$Fe.  The thin black line in the $f_{\rm
  L}$ panel at 570 MeV/$c$, is the empirical scaling function obtained
  by fitting the data.  The thick lines show the results of our
  calculations when all the effects beyond the RFG model have been
  considered (see text).  The full lines have been calculated for
  $^{12}$C, the dotted lines for $^{16}$O, and the dashed lines for
  $^{40}$Ca. The thin dashed lines show the RFG scaling functions.}
\label{fig:fexp}
\end{figure}

The other two kinds of scaling are not so well fulfilled by the
experimental functions. It is evident, from the figure, the poor
quality of the scaling of zeroth kind. Longitudinal and transverse
scaling functions are remarkably different, not only in size, but even
in their shapes. The excitation of subnucleonic degrees of freedom,
mainly the excitation of the $\Delta$ resonance, strongly affects
$f_{\rm T}$, while it is almost irrelevant in $f_{\rm L}$.  Also the
quality of the scaling of first kind is rather poor. These observations
are in agreement with those of Refs.  \cite{don99a,don99b,mai02} where
also data measured at large $q$ values have been used.

From the analysis of the scaling properties of the experimental
functions, we have extracted two benchmark values of $\cer$ and $\cde$
that we have used to gauge the quality of the scaling of our
theoretical functions. The values we have chosen are those related to
the $f_L$ functions at $q$=570 MeV/c, see Tab. \ref{tab:rdelta}, where
the quasi-elastic scattering mechanism works better.  In the
following, we shall consider that the scaling is violated when $\cer
>$0.096 or $\cde >$ 0.11. These numbers are obtained by adding the
uncertainty to the central benchmark values. The non scaling regions
will be indicated by the gray areas in the figures.

From the same set of data we extracted, see Appendix \ref{sec:expdata},
an empirical universal scaling function represented by the thin full
line in the lowest left panel of Fig. \ref{fig:fexp}.  This function,
which we called $f_{\rm U}^{\rm ex}$, is rather similar to the
universal empirical function given in Ref. \cite{mai02}.

We start now to consider the scaling of the theoretical functions.
The thick lines show the results of our calculations when various
effects beyond the RFG are introduced. These scaling functions have
been obtained by considering the nuclear finite size, the collective
excitations, the short-range correlations, the final state
interactions, and, in the case of the $f_{\rm T}$ functions, the
meson-exchange currents. 

The results presented by the thick lines have been obtained for three
different nuclei. The full lines represent the \car results, while the
dotted and dashed lines show, respectively, the results obtained for
\oxy and $^{40}$Ca.  The differences between these curves become
larger with decreasing $q$ values. We obtain $\cer$=0.03 and
$\cde$=0.05 for $f_{\rm L}$ at 570 MeV/$c$ and $\cer$=0.05 and
$\cde$=0.15 at 300 MeV/$c$.  The scaling of the $f_{\rm T}$ functions
is not as good. In this case we obtain $\cer$=0.03 and $\cde$=0.06 at
570 MeV/$c$ and $\cer$=0.10 and $\cde$=0.13 at 300 MeV/$c$. In any
case, these numbers are smaller than our empirical reference values,
and we can state that the scaling of second kind is satisfied.

On the contrary, the curves of Fig. \ref{fig:fexp} show a poor scaling
of first kind. The comparison between the $f_{\rm L}$ functions
calculated for the three $q$ values indicated in the figure, gives a
minimum $\cer$ value of 0.13, found for \car nucleus, and a maximum
value of 0.15, found for the \caI nucleus.  The minimum and maximum
values of the other index, $\cde$, are 0.18 and 0.30.  We found
similar, even if few percents larger, values also for the $f_{\rm T}$
functions.

The scaling of zeroth kind is rather well satisfied. By comparing the
$f_{\rm L}$ and $f_{\rm T}$ for each nucleus and each $q$ value we
found 0.04 as a maximum value of $\cer$. We found 0.11 for $\cde$,
slightly large even if below our empirical limiting value.  This
relatively large value of $\cde$ is due to the presence of sharp
resonances in the longitudinal and transverse responses at $q$=300
MeV/$c$ which appear at different excitation energies.  We have chosen
the longitudinal scaling function obtained for \oxy at $q=570$~MeV/$c$
as the theoretical universal scaling function that we called $f_{\rm
  U}^{\rm th}$.

In Fig. \ref{fig:fexp} the thin dashed lines show the RFG scaling
functions. It is evident that the inclusion of the effects beyond the
RFG we have considered, produce relevant modifications of
the RFG scaling functions. These modifications remarkably improve the
agreement with the experimental scaling functions. On the other hand,
the effects we have considered do not heavily modify the scaling
properties of the $f_{\rm L}$ and $f_{\rm T}$ functions. In the
remaining part of the section, we first discuss the consequences of
each effect beyond RFG, and then we analyze the scaling properties for
neutrino scattering processes.

\subsection{Finite size effects}
The starting point of our calculations is the continuum shell model.
In this model, the scattering processes are described by using some
assumption on the nuclear structure which are also used in the Fermi
gas model. We refer to the fact that both nuclear models consider the
nucleons free to move in a mean-field potential.  The continuum shell
model takes into account the finite dimensions of the system, the
finite number of nucleons, and the fact that protons and neutrons feel
different mean-field potentials.  In our calculations, the mean-field
is produced by a Woods-Saxon well.  The parameters of this potential
are taken from Refs. \cite{bot05} (for $^{12}$C) and \cite{ari96} (for
$^{16}$O and $^{40}$Ca) and have been fixed to reproduce the single
particle energies around the Fermi surface and the rms radii of the
charge distributions of each nucleus we have considered.

\begin{figure}
\begin{center}
\parbox[c]{16cm}
{\includegraphics[scale=0.5]{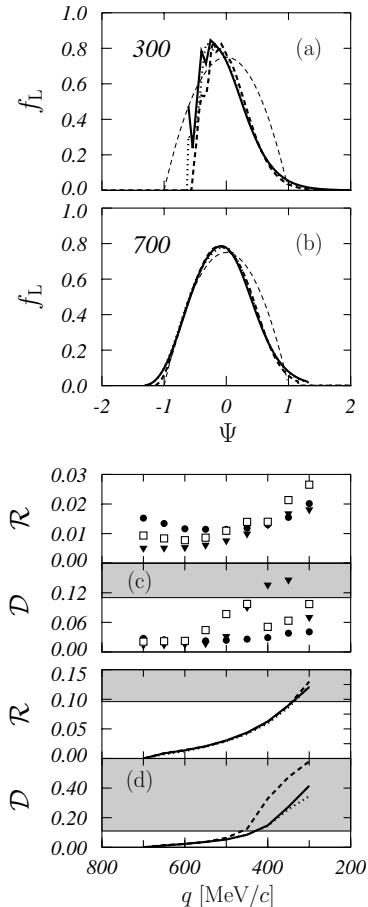}}
\end{center}
\vspace*{-0.8cm}
\caption{\small Continuum shell model results. In the panels (a) and
  (b), the thick lines represent the $f_{\rm L}$ scaling functions
  calculated for the various nuclei: full lines $^{12}$C, dotted lines
  $^{16}$O, dashed lines $^{40}$Ca.  The thin dashed lines represent
  the RFG scaling function.  The number inside the panels indicate the
  values of the momentum transfer in MeV/$c$ units.  In the panel (c)
  we show fr each nucleus the values of the indexes $\cer$ and $\cde$
  obtained at a fixed $q$ value by comparing the $f_{\rm L}$ and
  $f_{\rm T}$ functions.  The black circles indicate the \car results,
  the black triangles those of \oxy and the white squares those of
  $^{40}$Ca.  In the panel (d) we show the value of the two indexes
  obtained by considering the $f_L$ functions calculated for all the
  momentum transfer values ranging from the indicated $q$ value up to
  700 MeV/$c$. Details of the procedure are given in the text.  As in
  the panels (a) and (b), the full lines refer to $^{12}$C, the dotted
  ones to $^{16}$O, and the dashed ones to $^{40}$Ca.  The grey areas,
  drawn above the empirical values of $\cer$ and $\cde$, indicate the
  non-scaling region.  }
\label{fig:mfrel}
\end{figure}

The scaling properties of this model have been verified in Ref.
\cite{ama05b} for $q$ values larger than 700 MeV/$c$. We are
interested in the region of lower $q$ values, and we have calculated
longitudinal and transverse responses for momentum transfer values
down to 300 MeV/$c$. Our results are summarized in Fig.
\ref{fig:mfrel}.  In the (a) and (b) panels of the figure the thick
lines show the $f_{\rm L}$ scaling functions obtained respectively for
$q$=300 MeV/$c$ and $q$=700 MeV/$c$, the extreme values used in our
calculations. The full, dotted and dashed lines indicate the $^{12}$C,
\oxy \/ and \caI results, respectively, while the thin dashed lines
show the RFG model ones. As expected, for $q$=300 MeV/$c$, shell model
and RFG produce rather different scaling functions. The shell model
results present sharp resonances, and the figure indicates that the
scaling of second kind is poorly satisfied. The situation changes with
increasing momentum transfer. For $q$=700 MeV/$c$ the $f_{\rm L}$ show
excellent scaling of second kind and the agreement with the RFG
results has largely improved. Our scaling functions do not have their
maxima exactly at $\Psi$=0 and present a small left-right asymmetry.

A more concise information about the scaling properties of these
results, is given in the other two panels of Fig. \ref{fig:mfrel}.  In
the panel (c) the values of $\cer$ and $\cde$ are calculated by
comparing the 
$f_{\rm L}$ and $f_{\rm T}$ scaling functions
of the same nucleus, for a fixed
$q$ value.  The results shown in panel (c) give information how the
scaling of zeroth kind is verified at each $q$ value.  In this panel,
the black circles show the \car results, the black triangles the \oxy
results and the white squares the \caI results.  The general trend is
an increase of the indexes values at low $q$.  In any case, all the
values of the indexes shown in this panel are well below the empirical
ones, indicating the good quality of the scaling.

The results shown in the panel (d) have been obtained by using the
following procedure.  For each nucleus, we have calculated the scaling
functions from $q$=300 MeV/c up to $q$=700 MeV/c, in steps of 50
MeV/c.  The curves show the values of the indexes $\cer$ and $\cde$
obtained by considering in Eqs. (\ref{eq:delta}) and (\ref{eq:erre})
all the $f_L$ calculated from the $q$ value indicated in the figure,
up to $q$=700 MeV/c. Evidently, these curves are zero at $q$=700 MeV/c
and increase continuously with decreasing $q$ values.  The panel (d)
shows the evolution of the scaling of first kind with decreasing $q$
values.  In panel (d) the full lines show the \car results, the dotted
and dashed lines those of \oxy and of \caI respectively.  If the
scaling of first kind is verified, the values of $\cer$ and $\cde$ in
this panel are constant. We observe that all the curves are below the
empirical benchmark limits until the scaling function obtained for
$q$=400 MeV/$c$ is included. This could be considered the lower $q$
limit where the scaling of first kind is broken by the nuclear finite
size effects.

\subsection{Collective excitations}

By definition, mean-field models, such as the RFG model or the shell
model, do not describe collective excitations of the nucleus.  We have
considered the contribution of these excitations within a continuum
Random Phase Approximation (RPA) framework. Details of our RPA
calculations are given in Ref. \cite{bot05}. In the present work, we
used two effective nucleon-nucleon interactions.  They are a
zero-range interaction of Landau-Migdal type, called LM1 in
\cite{bot05}, and the finite-range polarization potential of
Ref. \cite{pin88}, properly renormalized as indicated in \cite{bot05},
and labeled PP.

Before discussing the results of the RPA calculations, we want to
point out a technical detail of our calculation. The semi-relativistic
prescription (\ref{eq:rele}) cannot be coherently implemented in the
continuum RPA equations. We have calculated the continuum RPA
responses without the semi-relativistic correction. The scaling
functions have been obtained from these responses by using the non
relativistic definition of the scaling variable \cite{dep98}:
\begin{equation}
\Psi= \frac {1}{k_{\rm F}} 
\left[\frac {m (\omega -E_{\rm shift})} {q} - \frac{q}{2}  \right] \, .
\end{equation}
With this scaling variable, the superscaling function of the  
non relativistic Fermi gas model assumes again the 
expression of Eq. (\ref{eq:scalFG}) (see Ref. \cite{dep98}).

\begin{figure}
\begin{center}
\parbox[c]{16cm}
{\includegraphics[scale=0.5]{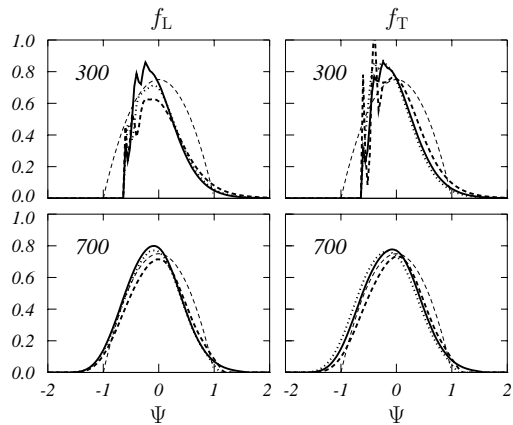}}
\end{center}
\vspace*{-0.8cm}
\caption{\small Scaling functions calculated for the \car nucleus.
  The thin dashed lines show the Fermi gas results.  The full lines
  show the mean-field results. The other lines have been obtained by
  using the continuum RPA. The thick dotted lines show the results
  obtained with the PP interaction, while the thick dashed lines have
  been obtained by using the LM1 interaction. The numbers in the
  panels indicate the values of the momentum transfer in MeV/$c$
  units.  }
\label{fig:srpa}
\end{figure}

The comparison between the Fermi gas scaling function and $f_{\rm L}$
and $f_{\rm T}$ calculated with the RPA is presented in
Fig. \ref{fig:srpa}.  In this figure, we show the \car scaling
functions obtained for the two extreme values of $q$ considered in our
calculations.  The thick full lines show the mean-field results, the
dotted lines the results obtained with the LM1 interaction, and the
dashed lines those obtained with the PP interaction.

The scaling functions at $q$=300 MeV/$c$ are strongly affected by the
RPA. This was expected, since for this value of the momentum transfer,
the maxima of the electromagnetic responses are very close to the
giant resonance region. The situation is rather different for the case
of 700 MeV/$c$ where the mean-field and RPA scaling functions are very
similar. Here, the RPA effects are larger for zero-range, than for
finite-range interaction.  The explanation of this fact becomes
evident if one considers the ring approximation of the RPA propagator
for a infinite system \cite{fet71}:
\begin{equation}
\Pi^{\rm RPA}(q,\omega) = 
\frac {\Pi^0(q,\omega)}{1-V(q)\Pi^0(q,\omega)} \, ,
\label{eq:ring}
\end{equation}
where $\Pi^0$ indicates the free Fermi gas polarization propagator,
and $V(q)$ is a purely scalar interaction. Finite-range interactions
vanish at large $q$ values, therefore the RPA propagator become equal
to that of the Fermi gas. This does not happen if contact interactions
are used, since these interactions are constant in momentum space. We
found that for $q$ values larger than 500 MeV/$c$, the RPA effects are
negligible if calculated with a finite-range interaction.

\begin{figure}[b]
\begin{center}
\parbox[c]{16cm}
{\includegraphics[scale=0.5]{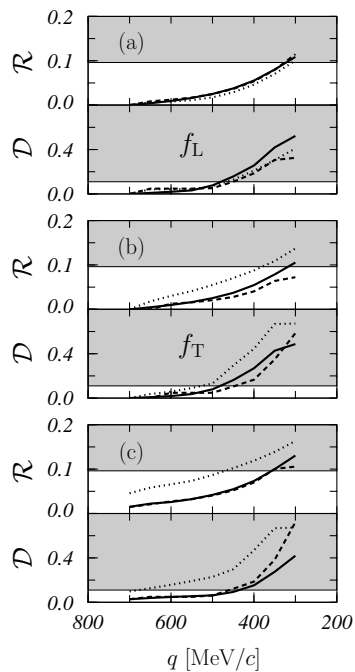}}
\end{center}
\vspace*{-0.8cm}
\caption{\small The $\cer$ and $\cde$ indexes calculated as in panel
  (d) of Fig.  \protect\ref{fig:mfrel}. In the panels (a) and (b) the
  $f_{\rm L}$ and $f_{\rm T}$ scaling functions calculated for the
  \car nucleus are separately shown.  In the panel (c) the indexes
  have been calculated by comparing the $f_{\rm L}$ and $f_{\rm T}$
  scaling functions.  The full lines represent the mean-field results,
  while the dotted and dashed lines have been obtained by doing
  continuum RPA calculations respectively with the polarization
  potential and with the Landau-Migdal interaction.}
\label{fig:irpac12}
\end{figure}

The scaling properties of continuum RPA $f_{\rm L}$ and $f_{\rm T}$
calculated for \car are summarized in Fig. \ref{fig:irpac12}. The
lines in this figure have been calculated with the same procedure used
in the panel (d) of Fig.  \ref{fig:mfrel}.  The full lines represent
the mean-field results, the dotted lines the results obtained with the
finite-range interaction and the dashes lines have been obtained with
the zero-range interaction.  The figure shows that the scaling of
first kind is well preserved by RPA calculations up to $q$=400
MeV/$c$. In the panel (c) the $f_{\rm L}$ and $f_{\rm T}$ scaling
functions have been put together in the calculation of the two
indexes.  We observe a worsening of the scaling, especially for the
polarization potential results. This indicates that the scaling of
zeroth kind is slightly ruined by the RPA. This is understandable,
since the effective nucleon-nucleon interaction acts in different
manner on the longitudinal and on the transverse nuclear responses.
Finally, the scaling of second kind is well preserved also in the RPA
calculations.

\subsection{Meson exchange currents}

We have seen that collective excitations are different in longitudinal
and transverse responses and this breaks the scaling of zeroth kind.
However, our RPA results show that these effects are too small to
explain the large differences between experimental $f_L$ and $f_T$
shown in Fig. \ref{fig:fexp}.  Another possible source of the breaking
of the zeroth kind scaling are the MEC.  Their role in the
longitudinal responses is negligible \cite{lal96}, while it can be
relevant in the transverse responses.

\begin{figure}[t]
\begin{center}
\parbox[c]{16cm}{\includegraphics[scale=.5]{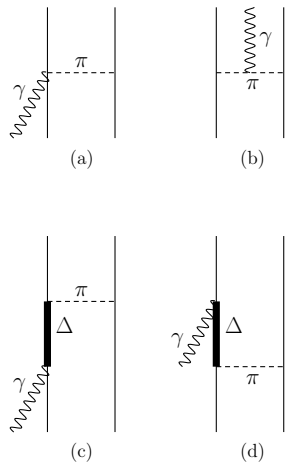}}
\end{center}
\vspace*{-0.7cm}
\caption{\small 
  Feynman diagrams of the MEC terms considered in our calculations.
  The (a) and (b) diagrams represent, respectively, the seagull and
  pionic currents, while the other two diagrams the $\Delta$ currents.
  }
\label{fig:diamec}
\end{figure}
\begin{figure}[t]
\begin{center}
\parbox[c]{16cm}
{\includegraphics[scale=0.5]{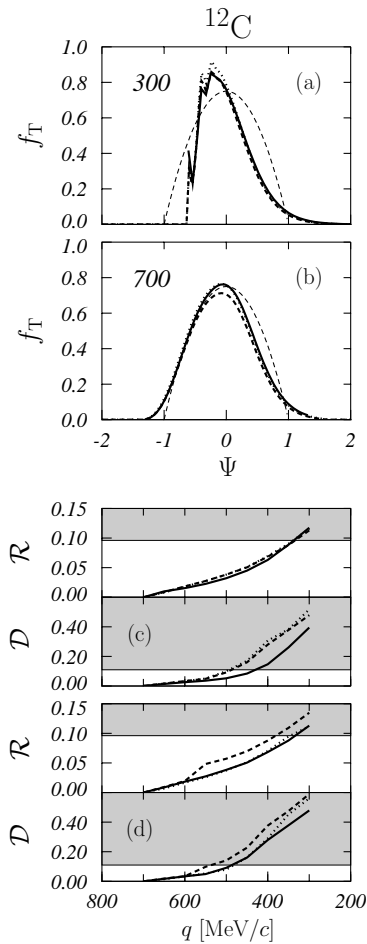}}
\end{center}
\vspace*{-0.8cm}
\caption{\small 
  Transverse scaling functions for the \car nucleus.  In the (a) and
  (b) panels, the thin dashed lines show the RFG model.  
  The other, thick,
  lines have been obtained by using the continuum shell
  model.  The full lines show the results obtained by using one-body
  currents only. The dotted lines have been obtained by inserting the
  pionic and seagull terms of the MEC, and the dashed lines show the
  results obtained by including also the $\Delta$ currents. The
  numbers inside the panels indicate the values of the momentum
  transfer in MeV/$c$.  The curves in the panels (c) and (d) are
  calculated as in panel (d) of Fig. \protect\ref{fig:mfrel}.  The
  curves in (c) compare the results obtained in \car by using one-body
  currents only (full line) with those obtained by adding seagull and
  pionic MEC (dotted line) and by adding also the $\Delta$ currents
  (dashed lines).  In panel (d) we compare the results obtained with
  all the MEC for the three nuclei considered. The full line show the
  \car result, the dotted line the \oxy result and the dashed line the
  \caI result.  
}
\label{fig:imec}
\end{figure}

We have calculated the transverse response functions by adding to the
one-body convection and magnetization currents the MEC arising
from the exchange of a single pion. In Fig. \ref{fig:diamec} we show
the Feynman diagrams of the MEC we have considered. They are the
seagull, or contact, term, represented by the (a) diagram of the
figure, where the virtual photon interacts at the pion-nucleon vertex,
and the pionic, or pion in flight term, represented by the (b)
diagram of the figure, where the virtual photon interacts with the
exchanged pion. In addition we consider also the $\Delta$ current
terms where the photon excites, or de-excites a virtual $\Delta$
resonance which interacts with another nucleon by exchanging a pion.
These $\Delta$ current terms are represented by the (c) and (d)
diagrams of Fig.  \ref{fig:diamec}.  A detailed description of our MEC
model is given in Refs. \cite{ang02,ang03,ang04}.

We show in panels (a) and (b) of Fig. \ref{fig:imec} the $f_{\rm T}$
scaling functions of the \car nucleus calculated for the extreme $q$
values we have considered.  The full lines have been obtained by using
one-body currents only, the dotted lines by including seagull and pionic
currents, and the dashed lines by adding the $\Delta$ currents. As
usual, the thin dashed lines show the RFG scaling function.

The effects of the MEC on the scaling functions are analogous to those
found on the responses in Ref. \cite{ama94}. The seagull and pionic
terms produce effects of opposite sign, therefore, the changes with
respect to the one-body responses are rather small, and almost vanish for
the largest values of $q$ we have considered. The inclusion of the
$\Delta$ currents slightly decreases the values of the scaling
functions. The presence of these currents becomes more relevant with
increasing $q$ value.

The panel (c) of Fig. \ref{fig:imec} shows the values of $\cer$ and
$\cde$, calculated for $f_{\rm T}$ as in panel (d) of
Fig. \ref{fig:mfrel}, for $^{12}$C. The meaning of the
different curves is the same as in the two upper panels of the
figure. In panel (d) we show the behaviour of the two indexes
calculated for the $f_{\rm T}$ scaling functions when all the MEC are
included. The full lines show the \car results, the dotted lines the
\oxy results and the dashed lines the \caI results.

Our MEC conserve rather well the scaling properties of $f_{\rm T}$.
The shapes of the $f_{\rm T}$, shown in the upper panels of
Fig. \ref{fig:imec}, are rather different from those of the empirical
$f_{\rm T}$, given in Fig.  \ref{fig:fexp}.  Our MEC model considers
only virtual excitations of the $\Delta$ resonance which become more
important at large $q$ values.  All these observations indicate that
the origin of the high energy tail of the experimental transverse
scaling functions is the real excitation of the $\Delta$ resonance,
with the consequent production of pions.

\subsection{Short-range correlations and final state interactions}

\begin{figure}
\begin{center}
\parbox[c]{16cm}
{\includegraphics[scale=0.4]{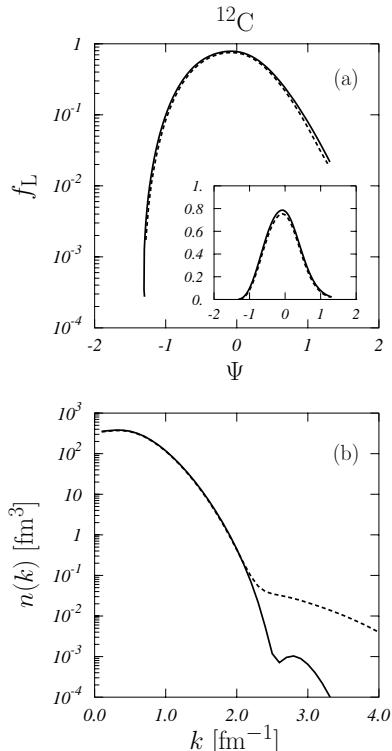}}
\end{center}
\vspace*{-0.8cm}
\caption{\small 
  In the panel (a) we show the longitudinal scaling function of \car
  calculated for $q$=700 MeV/$c$. The full lines show the mean-field
  result, the dashed lines have been obtained by including the SRC. In
  the insert, the same results are shown on a linear scale. In the panel
  (b) we show the momentum distribution of \car calculated with the
  mean-field model, full line, and with the SRC, dashed line.  }
\label{fig:src}
\end{figure}

As already mentioned, we have also investigated the influence of SRC.
Our results are summarized in Fig.  \ref{fig:src}, where, in the panel
(a), we show, as example, the $f_{\rm L}$ scaling function of \car
calculated for $q$=700 MeV/$c$ with (dashed lines) and without (full
curves) SRC.  The same curves are plotted in both linear and
logarithmic scales, to show the eventual effects in the tails of the
distributions.

\begin{figure}
\begin{center}
\parbox[c]{16cm}
{\includegraphics[scale=0.5]{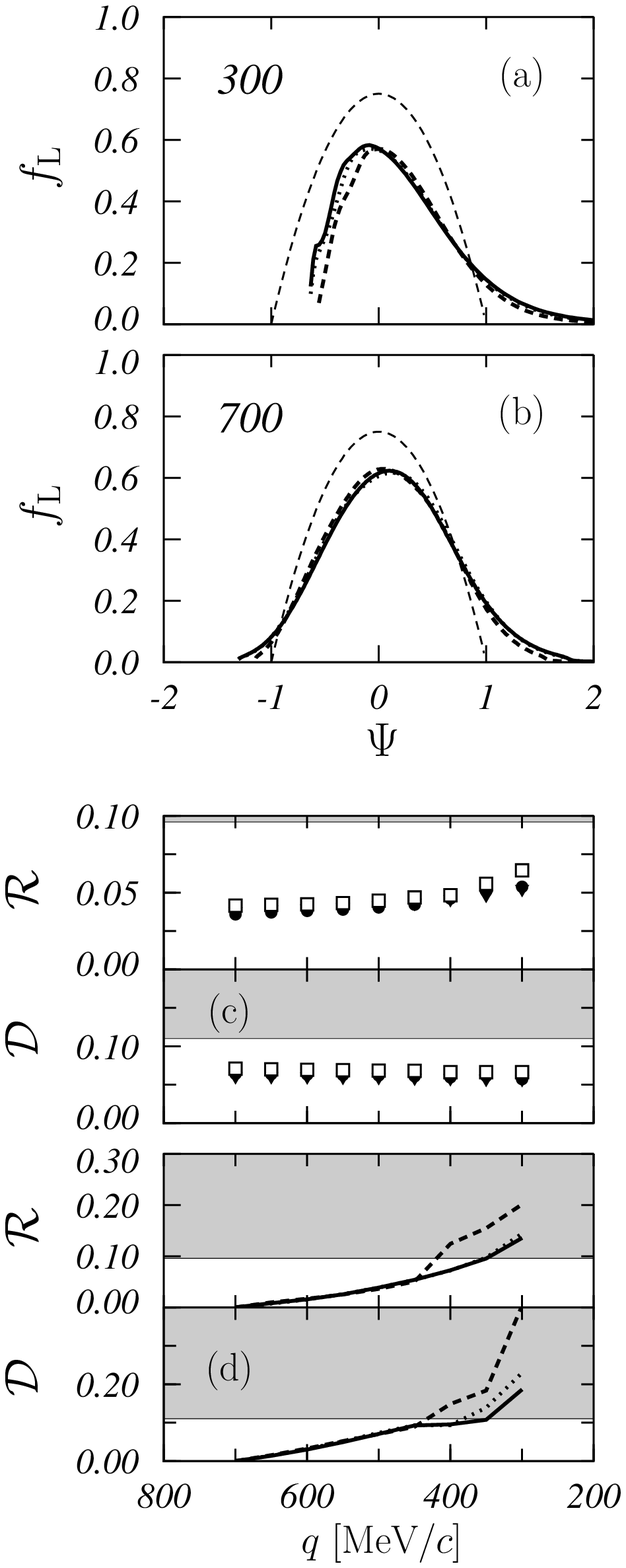}}
\end{center}
\vspace*{-0.8cm}
\caption{\small The same as Fig. \protect\ref{fig:mfrel} but showing
  the results of the mean field model with FSI.
}
\label{fig:ifold}
\end{figure}

The calculation of the responses with the SRC is done as described in
\cite{co01}, by considering all the cluster terms containing a single
correlation line. This implies the evaluation of two and tree points
cluster terms which produce contributions of different sign. The
calculations have been done with the scalar correlation function
labeled EU (Euler) in \cite{ari96}.  The effects of the correlations
on the momentum distribution of \car are shown in the panel (b) of
Fig. \ref{fig:src}. Also this momentum distribution has been
calculated in first order approximation \cite{ari97}. This example
show that the scaling functions, in the kinematics of our interest,
are insensitive to the high momentum tail of the momentum
distribution, and, in general, to the SRC.

All the results we have so far presented, did not include the FSI
which produce the largest modifications of the mean-field responses
\cite{ama01}. We treat the FSI by using the model developed in Refs.
\cite{co88,smi88}.  The mean-field responses are folded with Lorentz
functions whose parameters have been extracted from optical potential
volume integrals, and from the empirical spreading widths of single
particle states.  This approach has been successfully used to describe
quasi-elastic electromagnetic responses \cite{co88,ama94}, and, more
recently, it has been applied to calculate neutrino scattering cross
sections \cite{ble01,co02,bot05}.

In the two upper panels of Fig. \ref{fig:ifold} we show the shell
model $f_{\rm L}$ scaling functions corrected for the presence of the
FSI.  Again, we show here the results obtained for the two extreme
values of $q$ considered in our calculations.  The thin dashed lines
show the RFG results.  It is evident that the FSI are responsible for
the largest modifications of the mean-field results.  The values of
the maxima of the scaling functions in Fig.  \ref{fig:mfrel} are
around 0.8. After the inclusion of the FSI, these maxima are of the
order of 0.6.  The FSI lower the values of the maxima of the
responses, and, since the total area is conserved, increase their
widths.

The presence of the FSI slightly worsen the almost perfect scaling of
zeroth kind shown in Fig. \ref{fig:mfrel}.  The FSI act differently on
the two responses. The longitudinal responses are insensitive to the
spin and spin-isospin terms of the nuclear interaction. This fact is
considered in our FSI model. Even though in the panel (c) of Fig.
\ref{fig:ifold} the values of $\cer$, calculated for each $q$ value,
are slightly larger than the analogous ones of Fig. \ref{fig:mfrel},
they are below the empirical value.  The case of the $\cde$ index is
curious, since it shows almost constant values. This is because $\cde$
indicates the maximum difference between the various curves
considered. The FSI produce a smoothing of these curves and cancels
the sharp resonant peaks which appear at low $q$ values.

The curves in the panel (d) are obtained in the same way as those of
the analogous panel in Fig. \ref{fig:mfrel}.  The values shown in Fig.
\ref{fig:ifold} are clearly larger than those of panel (d) of
Fig. \ref{fig:mfrel}. For the $\cer$ index, the non scaling gray area
is reached when the $q$ value is about 500 MeV/$c$.  In conclusion, the
FSI produce large modifications of the mean-field responses, but do
not strongly violate the scaling.

\begin{figure}
\begin{center}
\parbox[c]{16cm}
{\includegraphics[scale=0.5]{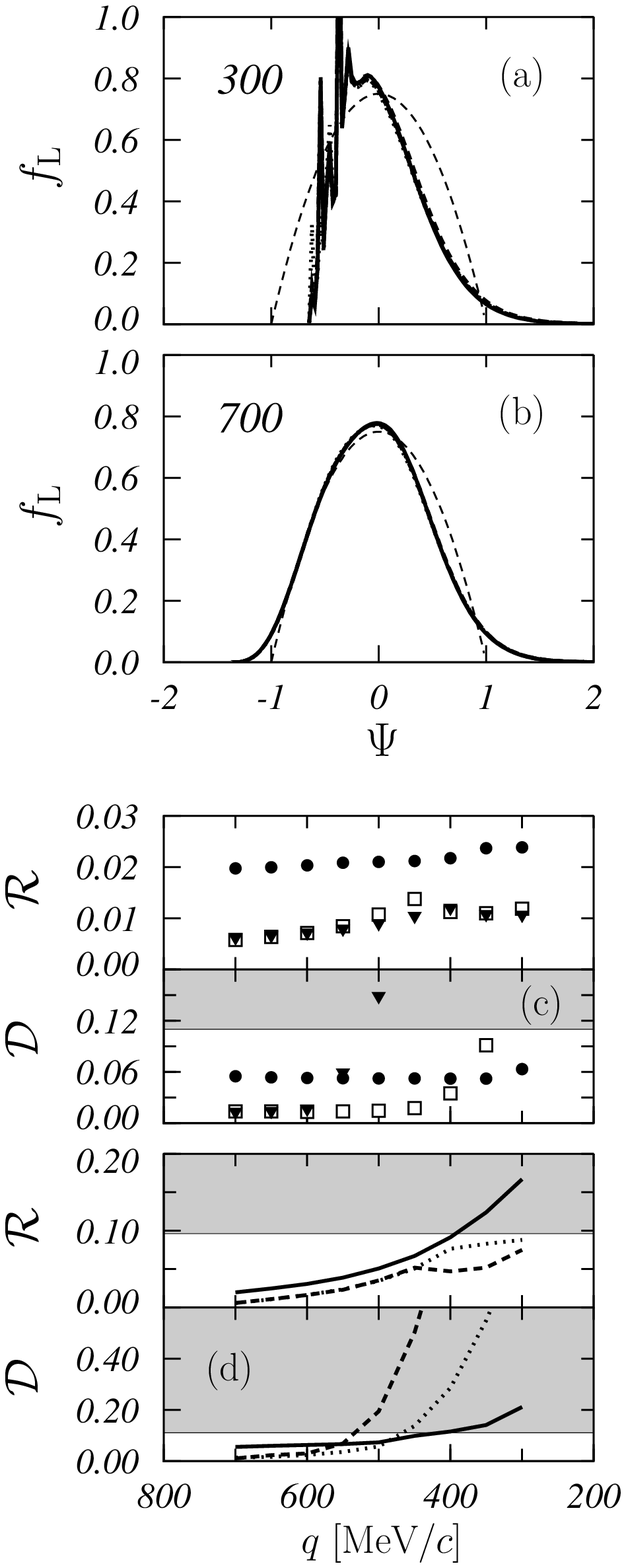}}
\end{center}
\vspace*{-0.8cm}
\caption{\small 
  The same as Fig. \protect\ref{fig:mfrel} for the neutrino scaling
  functions. In both panels (a) and (b) the five scaling functions
  defined in Eqs.  (\protect\ref{eq:fccv})-(\protect\ref{eq:ftpva}) and
  calculated for the \oxy nucleus are shown by the thick lines. These
  lines are almost exactly overlapped. Again the dashed thin lines
  show the RFG scaling functions.  In panel (c) the indexes are
  calculated by comparing the five scaling functions calculated at
  each $q$ value indicated on the $x$ axis.  The black dots show the
  \car results, the triangles the \oxy results and the white squares
  the \caI results.  The values the two indexes shown in panel (d)
  have been calculated as in the analogous panel of Fig.
  \protect\ref{fig:mfrel}.  The full line refers to $^{12}$C, the dotted
  one to \oxy and the dashed one to $^{40}$Ca.  
}
\label{fig:inu}
\end{figure}

\subsection{Neutrino scaling functions}

Up to now, we have discussed the scaling properties of the
electromagnetic scaling functions. We present in Fig. \ref{fig:inu}
the scaling functions defined in Eqs. (\ref{eq:fccv})-(\ref{eq:ftpva}),
for the $(\nu_e,e^-)$ charge exchange reaction.  The thick lines of the
two upper panels show the five scaling functions calculated, in a
continuum shell model, for the \oxy nucleus, and for the two extreme
values of $q$ considered in our work. The five curves are rather well
overlapped at $q$=300 MeV/$c$, and almost exactly overlapped at
$q$=700 MeV/$c$. The agreement with the RFG result, indicated as usual
by the dashed thin lines, is rather good at $q$=700 MeV/$c$.

In panel (c) we show the $\cer$ and $\cde$ indexes calculated by
comparing the five scaling functions at each $q$ value indicated in
the $x$ axis.  The black circles show the \car results, the black
triangles those of \oxy and the white squares the \caI results. These
values are of the same order of magnitude as those of the (c) panel of
Fig. \ref{fig:mfrel}.  This confirms the observation that the scaling
of zeroth kind is well satisfied in continuum shell model
calculations.
 
In panel (d) the values of the two indexes are evaluated by doing a
comparison of the scaling functions calculated at $q$=700 MeV/$c$ with
those obtained for lower $q$ values. This indicates the validity of
the scaling of first kind. The index $\cer$ shows that there is a
reasonable scaling down to $q$=400 MeV/$c$. This value is analogous to
that found for the electromagnetic functions.  The index $\cde$ shows
much rapid variations and, already at $q$=500 MeV/$c$, its value is over
the empirical limiting value. This is due to the presence of sharp
resonances at low $q$ values in some of the responses.

We studied the effects beyond the RFG for charge exchange neutrino
responses, by following the same steps used for the electromagnetic
responses.  To be precise, we did not calculate responses with SRC or
with the MEC, since from the results obtained for the electromagnetic
responses, we do not expect large changes of the mean field results
due to these effects.  We found effects of RPA and FSI analogous to
those of the electromagnetic case.

\section{SUPERSCALING PREDICTIONS}
\label{sec:res2}
In the previous section we have studied how the effects beyond the RFG
model modify the scaling function. We found that the main effects are
produced by the FSI. Despite the large modifications of the RFG
scaling functions, the scaling properties are not heavily destroyed.
For momentum transfer values above 500 MeV/$c$, our scaling functions
present values of the scaling indexes smaller than the empirical
benchmarks.  After having established the range of validity of the
superscaling hypothesis, we investigate, in this section, its
prediction power.  The strategy of our investigation consists in
comparing responses, and cross sections, calculated by using RPA, FSI
and eventually MEC and SRC, with those obtained by using our universal
scaling functions, both $f_{\rm U}^{\rm ex}$ and $f_{\rm U}^{\rm
  th}$.  All the RPA calculations presented in this section have been
done by using the PP interaction.

\begin{figure}
\begin{center}
\parbox[c]{16cm}
{\includegraphics[scale=0.48]{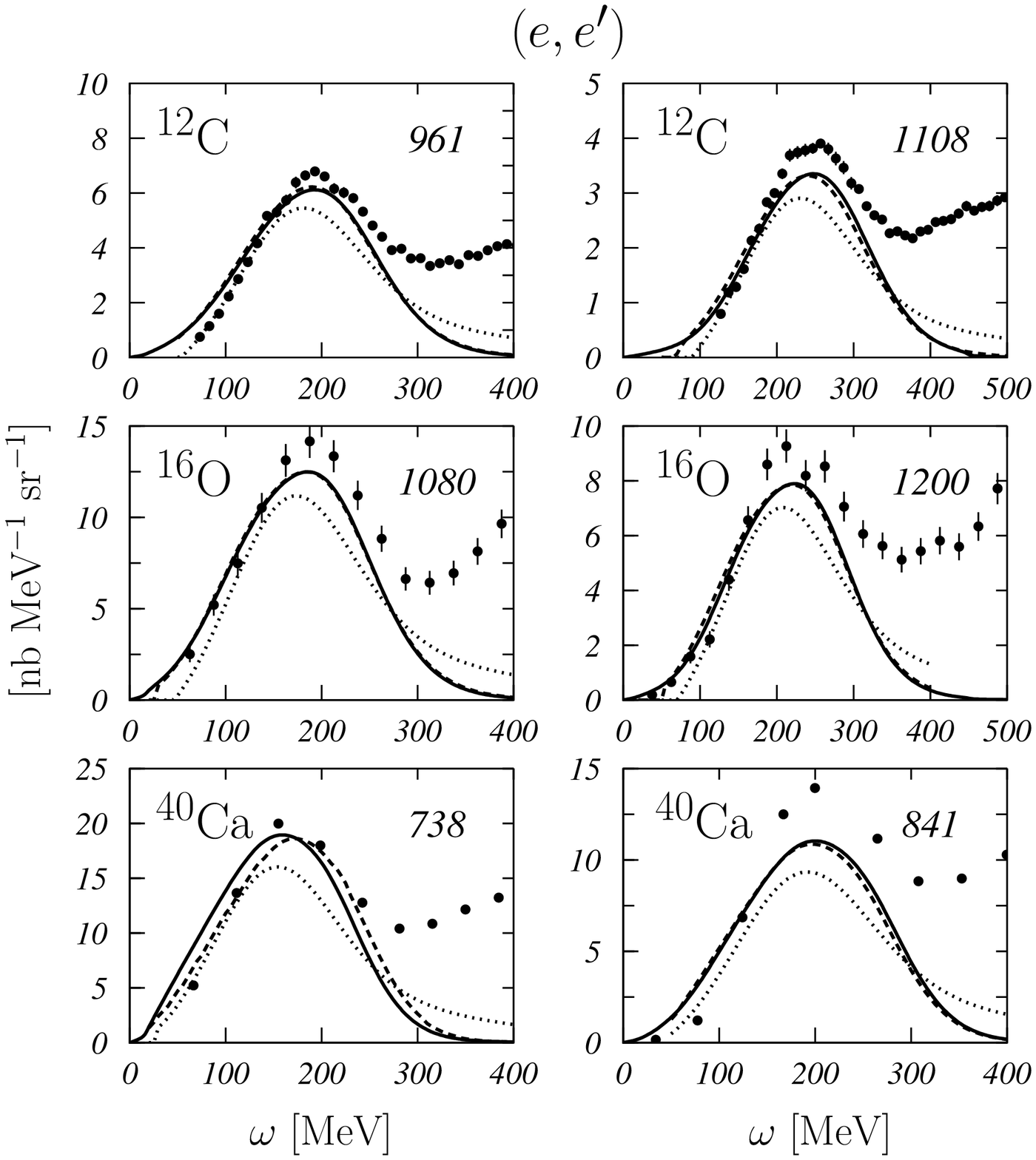}}
\end{center}
\vspace*{-0.8cm}
\caption{\small 
  Inclusive electron scattering cross sections.  Here, the numbers in
  the panels indicate, in MeV, the energy of the incoming
  electron.  The \car data \protect\cite{sea89} have been measured at
  a scattering angle of $\theta$=37.5$^o$, the \oxy data
  \protect\cite{ang96a} at $\theta$=32.0$^o$ and the \caI data
  \protect\cite{wil97} at $\theta$=45.5$^o$.  The full lines show the
  results of our complete calculations. The cross sections obtained by
  using $f_{\rm U}^{\rm th}$ are shown by the dashed lines,
  and those obtained with $f_{\rm U}^{\rm ex}$ are given by
  the dotted lines.  
}
\label{fig:ee_xsect}
\end{figure}
\begin{figure}
\begin{center}
\parbox[c]{16cm}
{\includegraphics[scale=0.45]{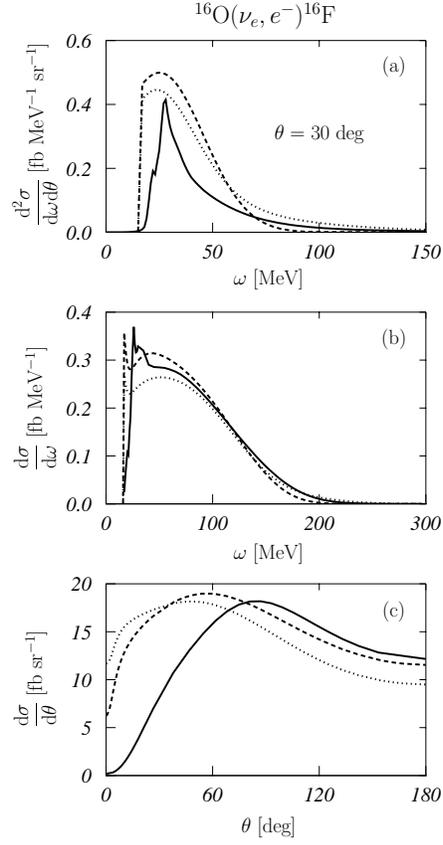}}
\end{center}
\vspace*{-0.8cm}
\caption{\small Neutrino charge exchange cross sections on
  $^{16}$O. All the results shown in the various panels have been
  obtained for neutrino energy of 300 MeV. In all the panels the full
  lines show the result of our complete calculation, while the other
  lines show the results obtained by using the scaling
  functions. Specifically, the dashed and the dotted lines have been
  obtained respectively with $f_{\rm U}^{\rm th}$ and $f_{\rm
  U}^{\rm ex}$. In the panel (a) the double differential cross
  sections calculated for the scattering angle of 30$^o$ as a function
  of the nuclear excitation energy is shown. In panel (b) we show the
  cross sections integrated on the scattering angle, always as a
  function of the nuclear excitation energy. In panel (c) we show the
  cross sections integrated on the nuclear excitation energy, as a
  function of the scattering angle.  }
\label{fig:nue}
\end{figure}
\begin{figure}
\begin{center}
\parbox[c]{16cm}
{\includegraphics[scale=0.35]{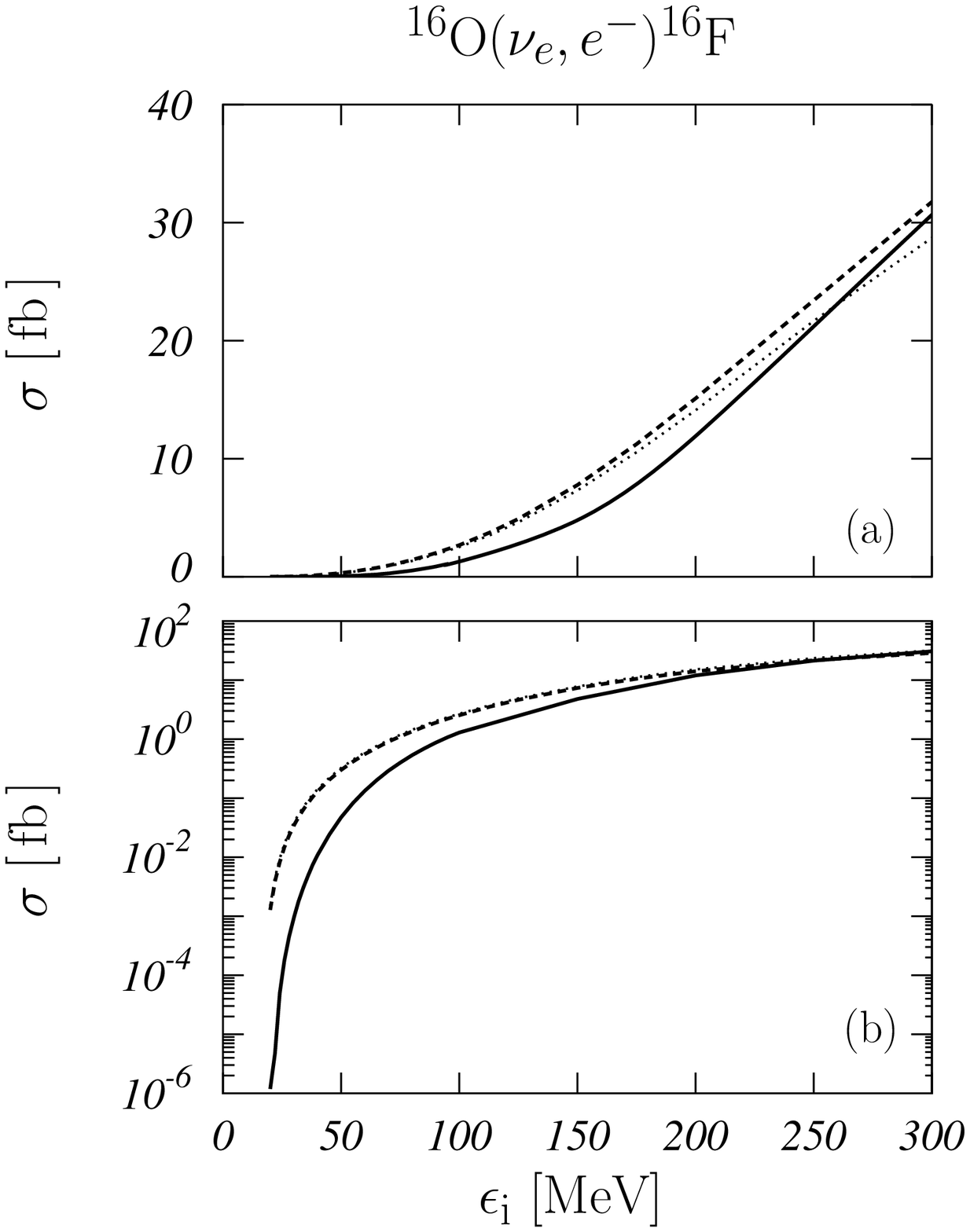}}
\end{center}
\vspace*{-0.8cm}
\caption{\small Total neutrino cross sections. Both panels show the
 same results in linear (a) and logarithmic (b) scales. The full lines
 show the result of the complete calculations. The dashed lines have
 been obtained by using $f_{\rm U}^{\rm th}$, and the dotted lines
 by using $f_{\rm U}^{\rm ex}$. }
\label{fig:totnue}
\end{figure}

The first test case of our study is done on the double differential
electron scattering cross section. We show in Fig. \ref{fig:ee_xsect}
the inclusive electron scattering cross sections calculated with our
model including the MEC and the FSI effects (full lines), those
obtained with $f_{\rm U}^{\rm th}$ (dashed lines) and the cross
sections obtained with $f_{\rm U}^{\rm ex}$  (dotted lines).
These results are compared with the data of Refs.
\cite{sea89,ang96a,wil97}.

The first remark about Fig. \ref{fig:ee_xsect}, regards the excellent
agreement between the results of the full calculations with those
obtained by using $f_{\rm U}^{\rm th}$. This clearly
indicates the validity of the scaling approach in this kinematic
region. This result was expected from the studies of the previous
section, since in all the cases shown in Fig. \ref{fig:ee_xsect}, the
value of the momentum transfer is larger than 500 MeV/$c$. The
differences with the cross sections obtained by using the empirical
scaling functions, reflect the differences between the various scaling
functions shown in Fig. \ref{fig:fexp}.

A second remark regarding Fig. \ref{fig:ee_xsect}, is about the fact
that our results underestimate the data. Probably this is because the
excitation of the $\Delta$ resonance is not considered in our
calculations.  The behaviour of the data of the figure in the higher
energy part, show the presence of the $\Delta$ resonance.  The low
energy tail of the excitation of this resonance affect also the
quasi-elastic peak.

The situation for the double differential cross sections is well
controlled, since all the kinematic variables, beam energy, scattering
angle, energy of the detected lepton, are precisely defined, and,
consequently, also energy and momentum transferred to the nucleus.
The situation changes for the total cross sections which are of major
interest for the neutrino physics. The total cross sections are only
function of the energy of the incoming lepton, therefore they consider
all the scattering angles and the possible values of the energy and
momentum transferred to the nucleus, with the only limitation of the
global energy and momentum conservations. This means that, in the
total cross sections, kinematic situations where the scaling is valid
and also where it is not valid are both present.

In order to clarify this point with quantitative examples, we show in
Fig.  \ref{fig:nue} various differential charge-exchange cross
sections obtained for 300 MeV neutrinos on \oxy target. In the panel
(a) we show the double differential cross sections calculated for a
scattering angle of 30$^o$, as a function of the nuclear excitation
energy.  The full line show the result of our complete calculation,
done with continuum RPA and FSI. We have shown in the
previous section that the effects of MEC and SRC are negligible, in
this kinematic regime. The dashed line show the result obtained with
$f_{\rm U}^{\rm th}$ and the dotted line with $f_{\rm U}^{\rm ex}$.
The values of the momentum transfer vary from about 150 to 200
MeV/$c$. Evidently this is not the quasi-elastic regime where the scaling
is supposed to hold, and this evidently produces the large differences
between the various cross sections.

The cross sections integrated on the scattering angle are shown as a
function of the nuclear excitation energy in the panel (b) of the
figure, while the cross sections integrated on the excitation energy
as a function of the scattering angle are shown in the panel (c). 

The three panels of the figure illustrate in different manners the same
physics issue. The calculation with the scaling functions fails in
reproducing the results of the full calculation in the region of low
energy and momentum transfer, where surface and collective effects are
important. This is shown in panel (b) by the bad agreement between the
three curves in the lower energy region, and in panel (c) at low
values of the scattering angle, where the $q$ values are minimal.
  
Total charge-exchange neutrino cross sections are shown in Fig.
\ref{fig:totnue} in both linear and logarithmic scale, as a function
of the energy of the incident neutrino $\epsilon_{\rm i}$. As in the
previous figure, the full lines show the result of the full
calculation, while the dashed and dotted lines have been respectively
obtained with $f_{\rm U}^{\rm th}$ and $f_{\rm U}^{\rm ex}$.  The
scaling predictions for neutrino energies up to 200 MeV are
unreliable. These total cross sections are obviously dominated by the
giant resonances, and more generally by collective nuclear excitation.
We have seen that these effects strongly violate the scaling. At
$\epsilon_{\rm i}$ =200 MeV the cross section obtained with $f_{\rm
U}^{\rm th}$ is about 20\% larger than those obtained with the full
calculation.  This difference becomes smaller with increasing energy
and is about 7\% at $\epsilon_{\rm i}$ = 300 MeV. This is an
indication that the relative weight of the non scaling kinematic
regions become smaller with the increasing neutrino energy.

\section{SUMMARY AND CONCLUSIONS}
\label{sec:conc}
We have investigated the scaling properties of the electron and
neutrino cross sections in a kinematic region involving momentum
transfer values smaller than 700 MeV/$c$. Since our working
methodology implies the numerical comparison of different scaling
functions, we defined two indexes, Eqs. (\ref{eq:delta}) and
(\ref{eq:erre}), to have a quantitative indication of the scaling
quality.

We have first analyzed the scaling properties of the experimental
electromagnetic responses given in Ref. \cite{jou96} for the $^{12}$C,
\caI and $^{56}$Fe nuclei.  We found the better scaling situation for
the longitudinal responses at 570 MeV/$c$. By considering these data
we obtained empirical values of the two indexes which we consider the
upper acceptable limit to have scaling. From a fit to the same set of
data we have also obtained an empirical scaling function, $f_{\rm
  U}^{\rm ex}$.

Our study of the role played by effects beyond the RFG model on the
scaling properties of the electroweak responses consisted in comparing
the values of the indexes obtained in our calculations with the
empirical values.  We found that finite size effects conserve the
scaling of first kind, the most likely violated, down to 400 MeV/$c$. We
have estimated the effects of the collective excitations by doing
continuum RPA calculations with two different residual
interactions. The RPA effects become smaller the larger is the value
of the momentum transfer.  At momentum transfer values above 600 MeV/$c$
the RPA effects are negligible if calculated with a finite-range
interaction, while zero-range interactions produce larger
effects. Collective excitations breaks scaling properties.  We found
that scaling of first kind is satisfied down to about 500 MeV/$c$.
 
The presence of the MEC violates the scaling of the transverse
responses.  From the quantitative point of view, MEC effects, at
relatively low $q$ values, are extremely small.  In our model, MEC
start to be relevant from $q \sim$ 600 MeV/$c$, especially these MEC
related to the virtual excitation of the $\Delta$ resonance. In our
calculations the real excitation of the $\Delta$ resonance, and the
consequent production of real pions, is not considered. Our nuclear
models deal with purely nucleonic degrees of freedom. Experimental
transverse responses, such as those shown in Fig. \ref{fig:fexp},
clearly show the presence of the $\Delta$ resonance peak, with
increasing value of the momentum transfer. Our model indicates 
that MEC do not destroy the scaling in the kinematic range of our
interest.

We have also investigated the effects of the SRC, which could also
violate the scaling. However, the size of these effects are so small
as to be negligible. The main modifications of the mean-field responses
are due to the FSI. When we applied the FSI we obtain, even for
$q$=700 MeV/$c$, scaling functions very different from those predicted
by the RFG model or by the mean field model, and rather similar to the
empirical one.  In any case also the FSI do not heavily break the
scaling properties.  We found that the scaling of first kind is
conserved down to $q$=450 MeV/$c$.

We have presented in detail only the results obtained for the
electromagnetic transverse and longitudinal responses since we found
for the weak responses, related to the neutrino scattering processes,
analogous results.  We can summarize the main points of this first
part of our investigation by saying that the effects beyond the RFG
model we have considered, strongly modify the scaling functions, but
do not destroy their scaling.  This explain the good scaling
properties of the experimental longitudinal electromagnetic responses,
which are not affected by the excitation of the $\Delta$ resonance, an
effect not included in our calculations.

After studying the scaling properties of the various responses we have
investigated the reliability of the cross sections predicted by using
the scaling functions. The idea is to assume that superscaling is
verified, i.e. all the three kinds of scaling we have considered, and
then to use the scaling functions to predict the cross sections.  The
cross sections calculated with our complete model have been compared
with those obtained by using as superscaling functions the empirical
scaling function fitting the 570 MeV longitudinal data of Ref.
\cite{jou96} $f_{\rm U}^{\rm ex}$, and our longitudinal scaling
function $f_{\rm U}^{\rm th}$.  We have chosen this last scaling
function as a theoretical universal scaling function.

We have verified that, in the quasi-elastic peak, the electron
scattering cross sections obtained with the full calculation are very
close to those obtained with $f_{\rm U}^{\rm th}$.  Also the
comparison with the data is rather good.  These calculations have been
done for momentum transfer values larger than 500 MeV/$c$, therefore
these results confirm the validity of the superscaling in the
quasi-elastic regime.  The problems arise in the evaluation of the
total neutrino cross sections.  In these cross sections, together with
the contribution of the quasi-elastic kinematics, where superscaling
is satisfied, there is also the contribution of kinematics regions
where there is not scaling.  We found that the scaling predictions of
the total neutrino cross sections are unreliable up to neutrino
energies of 200 MeV. At this point the scaling cross sections are 20\%
larger than those obtained by the full calculation. This difference
become smaller with increasing neutrino energy, and we found to be
reduced to about the 7\% at 300 MeV.  We stopped here our calculations
of the total cross section, since our model is not any reliable for
larger neutrino energies.  The comparison between double differential
cross sections calculated at excitation energies of 150 and 200 MeV,
for neutrino energies up to 1 GeV, gives an indication that the
difference between the total cross sections becomes smaller with
increasing neutrino energy.  It is worth pointing out, however, that
for neutrino energies larger than 300 MeV, the contribution of the
$\Delta$ resonance is not any more negligible, as we have implicitly
considered in our calculations.

\begin{acknowledgments}
This work has been partially supported by the agreement INFN-CICYT, by
the spanish DGI (FIS2005-03577) and by the MURST through the PRIN:
{\sl Teoria della struttura dei nuclei e della materia nucleare}.
\end{acknowledgments}

\appendix

\section{The experimental scaling functions}
\label{sec:expdata}

In this appendix we describe the procedure followed to obtain the
experimental Scaling Functions (SF), and also the empirical values of
the indexes $\cde$ and ${\cal R}$, from the electromagnetic response
data of Ref. \cite{jou96}.

The SF data have been obtained by inserting in Eqs. (\ref{eq:lscal})
and (\ref{eq:tscal}), the values of the experimental responses.  The
uncertainty on the SF data has been evaluated directly from the same
equations, taking into account the uncertainties of the original
response data.

The evaluation of $\cde$ and ${\cal R}$, Eqs.(\ref{eq:delta}) and
(\ref{eq:erre}), requires the knowledge of the various SF at the same
$\Psi$ points. Since the SF data are given for different values of
$\Psi$, we fixed a grid of $\Psi$ values, and we produced pseudo SF
data by doing a quadratic interpolation of the SF data previously
obtained.

The uncertainties of these pseudo SF data have been obtained by using
a Monte Carlo strategy.  Associated to each experimental SF point we
have generated a new point compatible with the Gaussian distribution
related to the experimental uncertainty. These new data formed a set
of SF points used to obtain pseudo data on the grid by quadratic
interpolation.  We repeated this procedure thousand times, and
obtained, for each value of $\Psi$ of the grid, a distribution of SF
points which allowed us to determine the corresponding uncertainty.

After having determined 
the uncertainties of the pseudo SF data, we
calculated the uncertainty of the $\cde$ index as:
\[
\sigma_{\cde} \, = \, 
\sqrt{[f^{\rm max}_i(\psi_{i_{\rm max}})]^2+
[f^{\rm min}_i(\psi_{i_{\rm max}})]^2} \, ,
\]
where $\Psi_{i_{\rm max}}$ is the value where the difference
$f^{\rm max}_i-f^{\rm min}_i$ 
reaches the  maximum value.

We have calculated the uncertainty on ${\cal R}$ in two steps. We
first evaluated the uncertainty of the sum
\[
S\, = \, \sum_{i=1,\ldots,K} \left[
f^{\rm max}_i-f^{\rm min}_i \right] \, ,
\]
in the numerator of Eq. (\ref{eq:erre}) by using a procedure
analogous to that used to obtain $\sigma_\cde$. That is,
\[
\sigma_{S} \, = \, 
\sqrt{
\sum_{i=1,\ldots,K} \left( [f^{\rm max}_i]^2
                   +[f^{\rm min}_i]^2  \right) } \, .
\]
To obtain the global uncertainty, we used again a Monte Carlo
strategy, and we calculated the ratio in Eq.  (\ref{eq:erre}) thousand
times by sampling the values of $S$ and of $f^{\rm max}$ within the
corresponding Gaussian distributions.

The empirical SF represented by the thin full line in
the $f_{\rm L}$ panel at 570 MeV/$c$ in Fig. \ref{fig:fexp}, has been
obtained as a best fit of all the experimental points shown in the panel.
The expression of our fitting function is:
\begin{equation}
f_{\rm U}^{\rm ex}(\Psi) \, = \,
\frac{A \exp(-\Psi^2) + B \Psi^2 + C \Psi + D }
{(\Psi+E)^2 + F^2}  \, .
\label{eq:empsf}
\end{equation}
with  $A$= 0.971, $B$=-0.067,
$C$= 0.385, $D$= 0.145, $E$= 0.366, $F$= 1.378. 

%
%
%

\end{document}